\documentclass[11pt,ams]{article}%
\usepackage{geometry}
\usepackage{amsmath}
\usepackage{amsfonts}
\usepackage{amssymb}
\usepackage{graphicx}%
\providecommand{\U}[1]{\protect\rule{.1in}{.1in}}
\begin{document}

\date{}
\title{\textbf{A study of the Higgs and confining phases in Euclidean $SU(2)$ Yang-Mills theories in $3d$ by taking into account the Gribov horizon }}
\author{\textbf{M.~A.~L.~Capri}$^{a}$\thanks{caprimarcio@gmail.com}\,\,,
\textbf{D.~Dudal}$^{b}$\thanks{david.dudal@ugent.be}\,\,,
\textbf{A.~J.~G\'{o}mez}$^{a}$\thanks{ajgomez@uerj.br}\,\,,
\textbf{M.~S.~Guimaraes}$^{a}$\thanks{msguimaraes@uerj.br}\,\,,
\textbf{I.~F.~Justo }$^{a}$\thanks{igorfjusto@gmail.com}\,\,,\\
\textbf{S.~P.~Sorella}$^{a}$\thanks{sorella@uerj.br}\ \thanks{Work supported by
FAPERJ, Funda{\c{c}}{\~{a}}o de Amparo {\`{a}} Pesquisa do Estado do Rio de
Janeiro, under the program \textit{Cientista do Nosso Estado}, E-26/101.578/2010.}\,\,\\[2mm]
{\small \textnormal{$^{a}$  \it Departamento de F\'{\i }sica Te\'{o}rica, Instituto de F\'{\i }sica, UERJ - Universidade do Estado do Rio de Janeiro,}}
 \\ \small \textnormal{\phantom{$^{a}$} \it Rua S\~{a}o Francisco Xavier 524, 20550-013 Maracan\~{a}, Rio de Janeiro, Brasil}\\
	 \small \textnormal{$^{b}$ \it Ghent University, Department of Physics and Astronomy, Krijgslaan 281-S9, 9000 Gent, Belgium}\normalsize}
%\author{\textbf{M.~A.~L.~Capri}$$\thanks{%
%primarcio@gmail.com}\,\,, \textbf{A.~J.~G\'{o}mez}\thanks{%
%ajgomez@uerj.br}\,\,, \textbf{M.~S.~Guimaraes}\thanks{%
%msguimaraes@uerj.br}\,\,, \textbf{S.~P.~Sorella}\thanks{%
%sorella@uerj.br}\, \thanks{%
%Work supported by FAPERJ, Funda{\c{c}}{\~{a}}o de Amparo {\`{a}} Pesquisa do
%Estado do Rio de Janeiro, under the program \textit{Cientista do Nosso Estado%
%}, E-26/101.578/2010.}\,\,,\,\,\\
%\textit{{\small { UERJ $-$ Universidade do Estado do Rio de Janeiro}}}%
%\\
%\textit{{\small {Instituto de F\'{\i}sica $-$ Departamento de F\'{\i}sica Te%
%\'{o}rica}}}\\
%\textit{{\small {Rua S{\~a}o Francisco Xavier 524, 20550-013 Maracan{\~a},
%Rio de Janeiro, RJ, Brasil.}}}}
\maketitle

\begin{abstract}
We study $SU(2)$ three-dimensional Yang-Mills theories in presence of Higgs fields in the light of the Gribov phenomenon. By restricting the domain of integration in the functional integral to the  first  Gribov horizon, we are able to discuss  a kind of  transition between the Higgs and the confining phase  in a semi-classical approximation. Both adjoint and fundamental representation for the Higgs field are considered, leading to a different phase structure.

\end{abstract}

\section{Introduction}

The understanding of the nonperturbative aspects of nonabelian gauge theories is one of the main challenging problems in quantum field theories. As an example, we may quote the transition between the confining and nonconfining phases of a Yang-Mills theory in presence of Higgs fields, see refs.\cite{Polyakov:1976fu,Cornwall:1998pt,Baulieu:2001vw} for analytical investigations and  \cite{Fradkin:1978dv,Nadkarni:1989na,Hart:1996ac,Caudy:2007sf,Maas:2010nc,Greensite:2011zz} for results obtained through numerical lattice simulations. \\\\In this work we aim at presenting a study of this topic by investigating $3d$ Yang-Mills theories in presence of Higgs fields. Both adjoint and fundamental representation for the Higgs field will be considered. As we  shall  see, they will give rise to a different phase structure. \\\\As the title of the paper let it understand, our investigation will be carried out from the perspective of the Gribov issue  \cite{Gribov:1977wm}, {\it i.e.} by taking into account the existence of the Gribov copies which are unavoidably present in the gauge-fixing procedure\footnote{See \cite{Sobreiro:2005ec,Vandersickel:2012tz} for a pedagogical introduction to the Gribov problem.}. In this framework, the dynamics of the model can be captured through the study of the two-point correlation functions of the gluon field. A Yukawa type behavior for the gluon propagator signals that the Higgs phase takes place, while a propagator of the Gribov type \cite{Gribov:1977wm,Sobreiro:2005ec,Vandersickel:2012tz} means that the theory lies in a confining phase. We remind here that a propagator of the Gribov type does not allow for a particle interpretation, as it exhibits complex poles. As such, it is suitable for the description of a confining phase. \\\\The phase structure of the theory turns out to be deeply related to the representation of the Higgs field. We shall first investigate the adjoint representation, the corresponding theory being known as the Georgi-Glashow model. For technical simplicity, we shall stick to the SU(2) gauge group here. In this case, we shall find that the third component of the gluon field, $A^3_\mu$, is always confined, for all values of the gauge coupling $g$ and of the {\it vev} $\nu$ of the Higgs field. In fact, the propagator $\langle A^3_\mu(q) A^3_\nu(-q) \rangle$ will be always of the Gribov type, displaying unphysical complex conjugate poles. The situation looks different for the off-diagonal components $A^{\alpha}_\mu, \alpha=1,2$. In the weak coupling regime, $g^2  \ll \nu^2$, the off-diagonal correlation function $\langle A^\alpha_\mu(q) A^\beta_\nu(-q) \rangle$ decomposes into the sum of two Yukawa propagators with real and positive masses. Though, only the heaviest mass component of this decomposition can be regarded as a physical mode, as the other component has a negative residue. According to \cite{Nadkarni:1989na}, this phase can be referred  to  as the $U(1)$ symmetric phase: the $A^3_\mu$ component is confined, while the off-diagonal, or charged, components $A^{\alpha}_\mu, \;\alpha=1,2$, exhibit a physical mode. This result is in agreement with Polyakov's seminal paper  \cite{Polyakov:1976fu} on the Georgi-Glashow model, where confinement of the $A^3_\mu$ component  at weak coupling is due  to the contribution of classical monopole solutions whose condensation gives rise to an area law for the Wilson loop.  Moreover, in the strong coupling regime, $g^2 \gg \nu^2$,  the poles of the off-diagonal components $A^\alpha_\mu$ become complex, the corresponding propagator being of the Gribov type. This signals that all components of the gauge field, $A^a_\mu, a=1,2,3$, are confined. Again, according to  \cite{Nadkarni:1989na}, this phase can be referred as the $SU(2)$ confining phase, in which all gauge modes are unphysical. Essentially, in the adjoint representation for the Higgs field, the theory exhibits two phases: the $U(1)$ symmetric phase at weak coupling, $g^2  \ll \nu^2$, in which only the $A^3_\mu$ mode is confined, and the $SU(2)$ confining phase at strong coupling,  $g^2 \gg \nu^2$, in which all modes get confined.  These results are in agreement with the numerical lattice simulations of  \cite{Nadkarni:1989na,Hart:1996ac}, for sufficiently large values of the quartic Higgs self-coupling. It worth noticing  that our results about the $A^3_\mu$ component ensure the absence of a massless state, a feature consistent with the nonperturbative dynamics of a super-renormalizable theory. In fact, according to  \cite{Jackiw:1980kv}, massless $3d$ super-renormalizable theories should develop a dynamical nonperturbative mass gap which prevents the appearance of infrared divergences which show up in the perturbative expansion.  For instance, in \cite{Cornwall:1998pt}, such a nonperturbative dynamical mass generation in the $3d$ Georgi-Glashow has been investigated by adding to the original action a non-polynomial gauge invariant mass term which induces vortex type solutions. In the present context, one might argue that the appearance of the Gribov mass parameters whose nonperturbative generation follows from  the restriction of the domain of integration to the first Gribov horizon, gives rise to a natural mechanism for the mass gap generation in $3d$ super-renormalizable nonabelian gauge theories, as already pointed out  in \cite{Dudal:2008rm}. In $4d$, all couplings are dimensionless and no a priori mass generation is required to ``protect'' the infrared region. One could thus expect quite some differences between the $3d$ and $4d$ Higgs theories, the latter is therefore presented elsewhere \cite{Capri:2012ah}.\\\\Things change considerably when the Higgs field is in the fundamental representation. In that case the gauge group $SU(2)$ is completely broken. At weak coupling,  $g^2  \ll \nu^2$, all propagators decompose into a sum of two Yukawa propagators with  positive masses. One of the components is unphysical due to a negative residue. However, the component with the largest mass is physical. Therefore, at weak coupling all gauge modes display a massive physical component. This is what can be called a Higgs phase. In the strong coupling, $g^2  \gg \nu^2$, the propagator of all gauge modes are of the Gribov type, exhibiting complex conjugate poles. This is the confining phase. Therefore, when the Higgs field is in the fundamental representation, we have a weak coupling Higgs phase and a strong coupling confining phase. Again, these results are in agreement with lattice numerical investigations \cite{Nadkarni:1989na,Hart:1996ac} at sufficiently large values of the Higgs quartic self-coupling. \\\\The paper is organized as follows. In Section 2, after a short introduction to the Georgi-Glashow model and to its  quantization in the Landau gauge, we discuss the implementation of the restriction to the Gribov region $\Omega$ in order to deal with the issue of the Gribov copies. This will be done by making use of the so called Gribov no-pole condition \cite{Gribov:1977wm,Sobreiro:2005ec,Vandersickel:2012tz}, which we shall adapt to the presence of the Higgs field. Further,  we solve the gap equations for the Gribov parameters and we analyze the ensuing behavior of the gluon two-point correlation functions. In Section 3 we address the  case in which the Higgs field is in the fundamental representation of $SU(2)$. In Section 4 we collect our conclusions.

\section{Higgs field in the adjoint representation. The Georgi-Glashow model and the restriction to the Gribov region $\Omega$}

The $SU(2)$ Georgi-Glashow model describes the interaction between gauge fields and a Higgs field $\Phi^a$ in the adjoint representation. Working in Euclidean space and adopting the Landau gauge, $\partial_\mu A^a_\mu=0$, the action of the model is specified by the following expression
\begin{equation}
S=\int d^{3}x\left(\frac{1}{4}F_{\mu \nu }^{a}F_{\mu \nu }^{a}+\frac{1}{2}%
D_{\mu }^{ab}\Phi ^{b}D_{\mu }^{ac}\Phi ^{c}+\frac{\lambda }{2}\left(
\Phi ^{a}\Phi ^{a}-\nu ^{2}\right) ^{2}+b^{a}\partial _{\mu }A_{\mu
}^{a}+\bar{c}^{a}\partial _{\mu }D_{\mu }^{ab}c^{b}\right)  \;, \label{S}
\end{equation}
where the covariant derivative is defined by
\begin{equation}
\left( D_{\mu }\Phi \right) ^{a}=\partial _{\mu }\Phi ^{a}+g\epsilon
^{abc}A_{\mu }^{b}\Phi ^{c} \;.
\end{equation}
The field $b^a$ stands for the Lagrange multiplier implementing the Landau gauge, $\partial_\mu A^a_\mu=0$, while $({\bar c}^a, c^a)$ are the Faddeev-Popov ghosts. The vacuum configuration which minimizes the energy is achieved by a constant scalar field satisfying
\begin{equation}
\Phi ^{a}\Phi ^{a}=\nu ^{2} \;.
\end{equation}
Setting
\begin{equation}
\left\langle \Phi ^{a}\right\rangle=\nu \delta ^{a3}\;, \label{higgsv}
\end{equation}
the Higgs field $\Phi^a$ can be decomposed as
\begin{eqnarray}  \label{A}
\Phi ^{a} &=&\varphi ^{a}+\nu \delta ^{a3} \;, \qquad \left\langle \varphi ^{a}\right\rangle=0 \;, \label{dec}
\end{eqnarray}
where $\varphi^a(x)$ parametrizes the fluctuations around the vacuum configuration.  \\\\Making use of the decomposition \eqref{dec}, for the quadratic part of the action involving the gauge field $A^a_\mu$,  one easily obtains
\begin{equation}
S_{quad}=\int d^{3}x\left( \frac{1}{4} { \left(  \partial_\mu A^a_\nu -\partial_\nu A^a_\mu  \right)} ^2 + b^a \partial_\mu A^a_\mu
+ \frac{g^{2}\nu ^{2}}{2}\left( A_{\mu }^{1}A_{\mu }^{1}+A_{\mu }^{2}A_{\mu
}^{2}\right)  \right)  \;, \label{quad}
\end{equation}
from which one could argue that the naive Higgs mechanism would take place, so that the off-diagonal components of the gluon field $A^{\alpha}_\mu, \; \alpha=1,2$, should acquire a mass $m_{H}^{2}=
g^{2}\nu ^{2}$, {\it i.e.}
\begin{equation}
\label{gluonoff}
\left\langle A_{\mu }^{\alpha }(p)A_{\nu }^{\beta }(-p)\right\rangle =\frac{\delta ^{\alpha \beta }}{p^{2}+m_{H}^{2}}\left( \delta _{\mu \nu }-%
\frac{p_{\mu }p_{\nu }}{p^{2}}\right) \;,
\end{equation}
while the third component $A_\mu^3$ should remain massless, namely
\begin{equation}
\left\langle A_{\mu }^{3}(p)A_{\nu }^{3}(-p)\right\rangle =\frac{1}{p^{2}}%
\left( \delta _{\mu \nu }-\frac{p_{\mu }p_{\nu }}{p^{2}}\right) \;. \label{zm}
\end{equation}
However, as was pointed out by Polyakov  \cite{Polyakov:1976fu}, the theory exhibits a different behavior. The action $(\ref{S})$ admits classical solitonic solutions, known as the 't Hooft-Polyakov monopoles\footnote{ These configurations are instantons in Euclidean space-time.} which play a relevant role in the dynamics of the model. In fact, it turns out that these configurations give rise to a monopole condensation at weak coupling, leading to a confinement of the third component $A^3_\mu$, rather than to a Higgs type behavior, eq.\eqref{zm}, a feature also confirmed by lattice numerical simulations  \cite{Nadkarni:1989na,Hart:1996ac}.\\\\As already mentioned in the Introduction, the aim of the present work is that of analyzing the nonperturbative dynamics of the Georgi-Glashow model by taking into account the Gribov copies. In the Landau gauge, this issue can be faced by restricting the domain of integration in the path integral to the so called Gribov region $\Omega$ \cite{Gribov:1977wm,Sobreiro:2005ec,Vandersickel:2012tz}, defined as the set of all transverse gauge configurations for which the Faddeev-Popov operator is strictly positive, namely
\begin{equation}
\Omega = \;\; \{ A^a_\mu\;, \; \partial_\mu A^a_\mu=0 \;, \; -\partial_\mu D^{ab}_\mu > 0\; \}   \label{gr}
\end{equation}
The region $\Omega$ is known to be convex and bounded in all directions in field space. The boundary of $\Omega$, where the first vanishing eigenvalue of the Faddeev-Popov operator appears, is called the first Gribov horizon. A way to implement the restriction to the region $\Omega$ has been worked out by Gribov in his original work. It amounts to impose the no-pole condition \cite{Gribov:1977wm,Sobreiro:2005ec,Vandersickel:2012tz} for the connected two-point ghost function $\mathcal{G}^{ab}(k;A) = \langle k | \left(  -\partial D^{ab}(A) \right)^{-1} |k\rangle $, which is nothing but the inverse of the Faddeev-popov operator $-\partial D^{ab}(A)$. One requires that  $\mathcal{G}^{ab}(k;A)$ has no poles at finite nonvanishing values of $k^2$, so that it stays always positive. In that way one ensures that the Gribov horizon is not crossed, {\it i.e.} one remains inside $\Omega$. The only allowed pole is at $k^2=0$, which has the meaning of approaching the boundary of the region $\Omega$.\\\\Here, it is worth noticing that monopoles configurations give rise to zero modes of the Faddeev-Popov operator \cite{Capri:2012ev,Maas:2005qt}, i.e.~they are located on the Gribov horizon. To some extent, this observation provides a
connection between Polyakov's results and those obtained by the restriction to the Gribov region. \\\\Following Gribov's procedure \cite{Gribov:1977wm,Sobreiro:2005ec,Vandersickel:2012tz}, for the connected two-point ghost function $\mathcal{G}^{ab}(k;A)$ at first order in the gauge fields,  one finds
\begin{equation}
\mathcal{G}^{ab}(k;A)=\frac{1}{k^{2}}\left( \delta ^{ab}-g^{2}\frac{k_{\mu
}k_{\nu }}{k^{2}}\int \frac{d^{3}q}{(2\pi )^{3}}\varepsilon
^{amc}\varepsilon ^{cnb}\frac{1}{(k-q)^{2}}\left( A_{\mu }^{m}(q)A_{\nu
}^{n}(-q)\right) \right)  \label{Ghost} \;,
\end{equation}%
where use has been made of the transversality condition $q_{\mu}A_{\mu}(q)=0$. \\\\In order to correctly take into account the presence of the Higgs vacuum, eq.\eqref{higgsv}, we decompose  $\mathcal{G}^{ab}(k;A)$ into diagonal and off-diagonal components, according to
\begin{equation}
\mathcal{G}^{ab}(k,A)=\left(
\begin{array}{cc}
\delta^{\alpha \beta}\mathcal{G}_{off}(k;A) & 0 \\
0 & \mathcal{G}_{diag}(k;A)
\end{array}
\right)
\end{equation}
where
\begin{eqnarray}
\mathcal{G}_{off}(k;A) &=&\frac{1}{k^{2}}\left( 1+g^{2}\frac{k_{\mu }k_{\nu }%
}{2k^{2}}\int \frac{d^{3}q}{(2\pi )^{3}}\frac{1}{(q-k)^{2}}\left( A_{\mu
}^{\alpha }(q)A_{\nu }^{\alpha }(-q)+2A_{\mu }^{3}(q)A_{\nu }^{3}(-q)\right)
\right)  \nonumber \\
&\equiv &\frac{1}{k^{2}}\left( 1+\sigma _{off}(k;A)\right)       \approx \frac{1}{k^{2}}\left( \frac{1}{1-\sigma
_{off}(k;A)}\right) \label{Goff} \;,  \\[5mm]
\mathcal{G}_{diag}(k;A) &=&\frac{1}{k^{2}}\left( 1+g^{2}\frac{k_{\mu }k_{\nu
}}{k^{2}}\int \frac{d^{3}q}{(2\pi )^{3}}\frac{1}{(q-k)^{2}}\left( A_{\mu
}^{\alpha }(q)A_{\nu }^{\alpha }(-q)\right) \right)  \nonumber \\
&\equiv &\frac{1}{k^{2}}\left( 1+\sigma _{diag}(k;A)\right)  \approx \frac{1}{k^{2}}\left( \frac{1}{1-\sigma
_{diag}(k;A)}\right) \label{Gdiag} \;.
\end{eqnarray}%
The quantities $\sigma_{off}(k;A), \; \sigma_{diag}(k;A)$ turn out to be  decreasing functions of the momentum $k$ \cite{Gribov:1977wm,Sobreiro:2005ec,Vandersickel:2012tz}. Thus, the no-pole condition for the ghost function $\mathcal{G}^{ab}(k,A)$ is implemented by imposing that \cite{Gribov:1977wm,Sobreiro:2005ec,Vandersickel:2012tz}
\begin{eqnarray}
\sigma _{off}(0;A) &\leq &1\;, \nonumber \\
\sigma _{diag}(0;A) &\leq &1   \label{np} \;,
\end{eqnarray}%
where $\sigma_{off}(0;A), \; \sigma_{diag}(0;A)$ are given by
\begin{eqnarray}
\sigma _{off}(0;A) &=&\frac{g^{2}}{3}\int \frac{d^{3}q}{(2\pi )^{3}}\frac{%
\left( A_{\mu }^{3}(q)A_{\mu }^{3}(-q)+\frac{1}{2}A_{\mu }^{\alpha
}(q)A_{\mu }^{\alpha }(-q)\right) }{q^{2}}  \;, \nonumber  \\
\sigma _{diag}(0;A) &=&\frac{g^{2}}{3}\int {\frac{d^{3}q}{(2\pi )^{3}}\frac{%
\left( A_{\mu }^{\alpha }(q)A_{\mu }^{\alpha }(-q)\right) }{q^{2}}   \label{sigma} \;.}
\end{eqnarray}
These expressions are obtained by taking the limit $k \rightarrow 0$ of eqs.\eqref{Goff},\eqref{Gdiag}, and by making use of the property
\begin{eqnarray}
A_{\mu }^{a}(q)A_{\nu }^{a}(-q) &=&\left( \delta _{\mu \nu }-\frac{q_{\mu }q_{\nu }%
}{q^{2}}\right) \omega (A)(q)   \nonumber \\
&\Rightarrow &\omega (A)(q)=\frac{1}{2}A_{\lambda }^{a}(q)A_{\lambda }^{a}(-q)
\end{eqnarray}
which follows from the transversality of the gauge field, $q_\mu A^a_\mu(q)=0$. Also, it is useful to remind that, for an arbitrary function $\mathcal{F}(p^2)$, we have
\begin{equation}
\int \frac{d^{3}p}{(2\pi )^{3}}\left( \delta _{\mu \nu }-\frac{%
p_{\mu }p_{\nu }}{p^{2}}\right) \mathcal{F}(p^2)=\mathcal{A}\;\delta _{\mu \nu }  \label{a}
\end{equation}%
where, upon contracting both sides of eq.\eqref{a} with $\delta_{\mu\nu}$,
\begin{equation}
\mathcal{A}=\frac{2}{3}\int \frac{d^{3}p}{(2\pi )^{3}}\mathcal{F}(p^2).
\end{equation}%

\subsection{Gribov's  gap equations}

In order to implement the restriction to the Gribov region $\Omega$ in the functional integral, we encode the information of the no-pole conditions into step functions \cite{Gribov:1977wm,Sobreiro:2005ec,Vandersickel:2012tz}:
\begin{equation}
Z=\int {[DA_{\mu }]\delta (\partial A)(\det\mathcal{M})\theta (1-\sigma
_{diag}(A))\theta (1-\sigma _{off}(A))e^{-S_{YM}}.}
\end{equation}
Though, being interested in the study of the gluon propagators, we shall consider the quadratic approximation for the partition function, namely  \begin{eqnarray}
Z_{quad} &=&\int \frac{d\beta }{2\pi i\beta }\frac{d\omega }{2\pi i\omega }%
DA_{\mu }e^{\beta (1-\sigma _{diag}(0,A))}e^{\omega \left( 1-\sigma
_{off}(0,A)\right) }  \nonumber  \label{Zq} \\
&\times &e^{-\frac{1}{4}\int d^{3}x(\partial _{\mu }A_{\nu }^{a}-\partial
_{\nu }A_{\mu }^{a})^{2}-\frac{1}{2\xi }\int {d^{3}x(\partial _{\mu }A_{\mu
}^{a})^{2}-}\frac{{g^{2}\nu ^{2}}}{2}\int {d^{3}xA_{\mu }^{\alpha }A_{\mu
}^{\alpha }}} \;,
\end{eqnarray}
where use has been made of the integral representation
\begin{equation}
\theta(x) = \int_{-i \infty +\epsilon}^{i\infty +\epsilon} \frac{d\beta}{2\pi i \beta} \; e^{\beta x}  \;. \label{step}
\end{equation}
After simple algebraic manipulations, one gets
\begin{equation}
Z_{quad}=\int \frac{d\beta e^{\beta }}{2\pi i\beta }\frac{d\omega e^{\omega }%
}{2\pi i\omega }DA_{\mu }^{\alpha }DA_{\mu }^{3}e^{-\frac{1}{2}\int \frac{%
d^{3}q}{(2\pi )^{3}}A_{\mu }^{\alpha }(q)\mathcal{P}_{\mu \nu }^{\alpha
\beta }A_{\nu }^{\beta }(-q)-\frac{1}{2}\int \frac{d^{3}q}{(2\pi )^{3}}%
A_{\mu }^{3}(q)\mathcal{Q}_{\mu \nu }A_{\nu }^{3}(-q)},  \label{Zq1}
\end{equation}
with
\begin{eqnarray}
\mathcal{P}_{\mu \nu }^{\alpha \beta } &=&\delta ^{\alpha \beta }\left(
\delta _{\mu \nu }\left( q^{2}+\nu ^{2}g^{2}\right) +\left( \frac{1}{\xi }%
-1\right) q_{\mu }q_{\nu }+2\frac{g^{2}}{3}\left( \beta +\frac{\omega }{2}%
\right) \frac{1}{q^{2}}\delta _{\mu \nu }\right)  \label{P}  \;, \nonumber \\
\mathcal{Q}_{\mu \nu } &=&\delta _{\mu \nu }\left( q^{2}-2\frac{\omega g^{2}%
}{3}\frac{1}{q^{2}}\right) +\left( \frac{1}{\xi }-1\right) q_{\mu }q_{\nu } \;.
\label{Q}
\end{eqnarray}
The parameter $\xi$ stands for the usual gauge fixing parameter, to be set to zero at the end in order to recover the
Landau gauge. Evaluating  the inverse of the
expressions above and taking the limit $\xi\rightarrow 0$, for the gluon propagators one gets
\begin{eqnarray}
\left\langle A_{\mu }^{3}(q)A_{\nu }^{3}(-q)\right\rangle &=&\frac{q^{2}}{%
q^{4}+\frac{2\omega g^{2}}{3}}\left( \delta _{\mu \nu }-\frac{q_{\mu }q_{\nu
}}{q^{2}}\right)  \label{Pdiag} \;, \\
\left\langle A_{\mu }^{\alpha }(q)A_{\nu }^{\beta }(-q)\right\rangle
&=&\delta ^{\alpha \beta }\frac{q^{2}}{q^{2}\left( q^{2}+g^{2}\nu^{2}\right)
+g^{2}\left( \frac{2\beta }{3}+\frac{\omega }{3}\right) }\left( \delta _{\mu
\nu }-\frac{q_{\mu }q_{\nu }}{q^{2}}\right)  \label{NPoff} \;.
\end{eqnarray}
It remains to find the gap equations for the Gribov parameters $(\beta, \omega)$, enabling us to express them in terms of the parameters of the starting model, {\it i.e.} the gauge coupling constant $g$ and the {\it vev} of the Higgs field $\nu$. In order to accomplish this task we follow \cite{Gribov:1977wm,Sobreiro:2005ec,Vandersickel:2012tz} and evaluate the partition function $Z_{quad}$ in the semiclassical approximation. First, we integrate out the gauge fields, obtaining
\begin{equation}
Z_{quad}=\int{\frac{d\beta}{2\pi i\beta}\frac{d\omega}{2\pi i\omega}}%
e^{\beta}e^{\omega}\left(\det\mathcal{Q}_{\mu\nu}\right)^{-\frac{1}{2}%
}\left(\det\mathcal{P}^{\alpha\beta}_{\mu\nu}\right)^{-\frac{1}{2}} \;.
\label{Zq2}
\end{equation}
Making use of
\begin{equation}
\left(\det \mathcal{A}_{\mu\nu}^{ab}\right)^{-\frac{1}{2}}=e^{-\frac{1}{2}%
\ln \det \mathcal{A}_{\mu\nu}^{ab}}=e^{-\frac{1}{2}Tr \ln \mathcal{A}%
_{\mu\nu}^{ab}} \;,
\end{equation}
for the determinants in expression (\ref{Zq2}) we get
\begin{eqnarray}
\left( \det \mathcal{Q}_{\mu \nu }\right) ^{-\frac{1}{2}} &=&\exp \left[
-\int {\frac{d^{3}q}{(2\pi )^{3}}\ln \left( q^{2}+\frac{2\omega g^{2}}{3}%
\frac{1}{q^{2}}\right) }\right] \;, \nonumber \\
\left( \det \mathcal{P}_{\mu \nu }^{\alpha \beta }\right) ^{-\frac{1}{2}}
&=&\exp \left[ -2\int {\frac{d^{3}q}{(2\pi )^{3}}\ln \left(
(q^{2}+g^{2}\nu^{2})+g^{2}\left( \frac{2\beta }{3}+\frac{\omega }{3}\right)
\frac{1}{q^{2}}\right) }\right] \;.
\end{eqnarray}
Therefore,
\begin{equation}  \label{Zq3}
Z_{quad}=\int{\frac{d\beta}{2\pi i}\frac{d\omega}{2\pi i}}e^{f(\omega,\beta)} \;,
\end{equation}
with
\begin{eqnarray}
f(\omega ,\beta ) &=&\beta +\omega -\ln \beta -\ln \omega -\int {\frac{d^{3}q%
}{(2\pi )^{3}}\ln \left( q^{2}+\frac{2\omega g^{2}}{3}\frac{1}{q^{2}}\right)
} \nonumber  \\
&-&2\int {\frac{d^{3}q}{(2\pi )^{3}}\ln \left(
(q^{2}+g^{2}\nu^{2})+g^{2}\left( \frac{2\beta }{3}+\frac{\omega }{3}\right)
\frac{1}{q^{2}}\right) }
\end{eqnarray}
Expression (\ref{Zq3}) can be now evaluated in the saddle point approximation \cite{Gribov:1977wm,Sobreiro:2005ec,Vandersickel:2012tz}, {\it i.e.}
\begin{equation}
Z_{quad}\approx e^{f(\beta^*,\omega^*)} \;,
\end{equation}
where the parameters $\beta^*$ and $\omega^*$ are determined by the stationary conditions
\begin{equation}
\frac{\partial f}{\partial \beta^*}=\frac{\partial f}{\partial \omega^*}=0 \;,
\end{equation}
which yield the following gap equations\footnote{We remind here that the terms $\log\beta$ and $\log \omega$ can be neglected in the derivation of the gap equations, eqs.\eqref{gap1} \eqref{gap2}, when taking the thermodynamic limit, see \cite{Gribov:1977wm,Sobreiro:2005ec,Vandersickel:2012tz} for details.}:
\begin{eqnarray}
\frac{g^{2}}{2}\int \frac{d^{3}q}{(2\pi )^{3}}\left( \frac{1}{q^{4}+\frac{%
2\omega ^{\ast }g^{2}}{3}}+\frac{1}{q^{2}(q^{2}+g^{2}\nu^{2})+g^{2}\left(
\frac{2\beta ^{\ast }}{3}+\frac{\omega ^{\ast }}{3}\right) }\right) &=&1 \;, \label{gap1} \\
\frac{g^{2}}{2}2\int \frac{d^{3}q}{(2\pi )^{3}}\left( \frac{1}{%
q^{2}(q^{2}+g^{2}\nu^{2})+g^{2}\left( \frac{2\beta ^{\ast }}{3}+\frac{\omega
^{\ast }}{3}\right) }\right) &=&1 \;, \label{gap2}
\end{eqnarray}
allowing us to express $\beta^*, \omega^*$ in terms of the parameters $\nu,g$.\\\\Equations \eqref{gap1} and \eqref{gap2} can be rewritten as
\begin{eqnarray}
2 \left( \frac{g^{2}}{2}\right) \int \frac{d^{3}q}{(2\pi )^{3}}\left( \frac{1}{%
q^{4}+\omega^* \frac{2g^{2}}{3}}\right) &=&1\;, \label{ggap1} \\
2 \left( \frac{g^{2}}{2}\right) \int \frac{d^{3}q}{(2\pi )^{3}}\left( \frac{1%
}{q^{2}(q^{2}+g^{2}\nu^{2})+(\beta^* \frac{2g^{2}}{3}+\omega^* \frac{g^{2}}{3})}
\right) &=&1\label{ggap2} \;.
\end{eqnarray}
The first integral is easy to compute, giving $\omega^* $ as a function of $g$
only
\begin{eqnarray}
\omega^*(g)= \frac{3}{2^{11}\pi^4} \;g^6        \;.  \label{omega}
\end{eqnarray}
In order to solve the second gap equation, eq.\eqref{ggap2},  we compute the roots of the
denominator:
\begin{equation}
q_{\pm }^{2}=\frac{-g^{2}\nu^{2}\pm \sqrt{g^{4}\nu^{4}-4\tau }}{2}  \label{p+-}  \;.
\end{equation}%
Notice that the roots are real when
\begin{equation}
 \tau \leq \frac{g^{4}\nu ^{4}}{4}  \label{cond}
\end{equation}%
with
\begin{equation}
\tau =\beta^* \frac{2g^{2}}{3}+\omega^* \frac{g^{2}}{3}  \label{tau}  \;.
\end{equation}
After decomposition in partial fractions,  equation \eqref{ggap2}  becomes
\begin{eqnarray}
\frac{4g^{2}\pi }{(2\pi )^{3}(q_{+}^{2}-q_{-}^{2})}\int_{0}^{\infty
}dq\left( \frac{q^{2}}{(q^{2}-q_{+}^{2})}-\frac{q^{2}}{(q^{2}-q_{-}^{2})}%
\right) &=&1 \;.
\end{eqnarray}%
Using the principal value prescription, this yields the final (finite) result
\begin{equation}
\frac{ig^{2}}{(4\pi )}\frac{1}{q_{+}-q_{-}}=1 \;.  \label{one}
\end{equation}
Making use of expression \eqref{p+-}, equation \eqref{one} gives $\tau$ as function of the parameters $(\nu,g)$, {\it i.e.}
\begin{equation}
\tau =\beta^* \frac{2g^{2}}{3}+\omega^* \frac{g^{2}}{3}  = \left[ \frac{1}{2}g^{2}\nu^{2}-\frac{g^{4}}{32\pi^2 }\right] ^{2}  \;. \label{feq}
\end{equation}
\subsection{Gluon propagators and phases of the theory}
Having solved the gap equations, we can now look at the behavior of the gluon propagators and discuss the phases of the theory. Let us start by looking at the propagator of the third component $A^3_\mu$, namely
\begin{equation}
\left\langle A_{\mu }^{3}(q)A_{\nu }^{3}(-q)\right\rangle =\frac{q^{2}}{%
q^{4}+\frac{2\omega^* g^{2}}{3}}\left( \delta _{\mu \nu }-\frac{q_{\mu }q_{\nu
}}{q^{2}}\right)  \label{Pdiagf} \;, \qquad \omega^*(g)= \frac{3}{2^{11}\pi^4} \;g^6  \;.
\end{equation}
One observes that expression \eqref{Pdiagf} turns out to be independent from the $vev$ $\nu$ of the Higgs field. It is of the Gribov type, displaying two complex conjugate poles. In other words, the mode $A^3_\mu$ is always confined, for all values of the parameters $g, \nu$. Concerning now the off-diagonal gluon propagator, it can be decomposed into the sum of two Yukawa modes, {\it i.e.}
\begin{eqnarray}
\left\langle A_{\mu }^{\alpha }(q)A_{\nu }^{\beta }(-q)\right\rangle
&=&\delta ^{\alpha \beta }\frac{q^{2}}{q^{2}\left( q^{2}+g^{2}\nu^{2}\right)
+g^{2}\left( \frac{2\beta^* }{3}+\frac{\omega^* }{3}\right) }\left( \delta _{\mu
\nu }-\frac{q_{\mu }q_{\nu }}{q^{2}}\right)  \nonumber  \\
&=& \delta ^{\alpha \beta }  \left(  \frac{{\cal R}_{+}}{q^2+m^2_+} -\frac{{\cal R}_{-}}{q^2+m^2_-}  \right)
\left( \delta _{\mu\nu }-\frac{q_{\mu }q_{\nu }}{q^{2}}\right)  \;,  \label{NPoff_f}
\end{eqnarray}
with
\begin{equation}
m^2_+ = \frac{g^2\nu^2 + \sqrt{g^4 \nu^4 - 4 \tau}}{2}  \;,  \qquad m^2_- = \frac{g^2\nu^2 - \sqrt{g^4 \nu^4 - 4 \tau}}{2} \label{masses} \;,
\end{equation}
and
\begin{equation}
{\cal R}_{+} = \frac{m^2_+}{m^2_+-m^2_-}  \;, \qquad {\cal R}_{-} = \frac{m^2_-}{m^2_+-m^2_-}  \label{residues} \;.
\end{equation}
We see thus that, when $\tau < \frac{g^2\nu^2}{4}$, both masses $m^2_+, m^2_-$ are real, positive and  different, as well as the two quantities ${\cal R}_{+}, {\cal R}_{-}$. Moreover, due to the presence of the relative minus sign in expression \eqref{NPoff_f}, only the heaviest mode with mass $m^2_+$ represents a physical excitation. \\\\Therefore, taking into account eq.\eqref{feq}, we can identify the following regions in the $(g^2,\nu^2)$ plane:
\begin{itemize}
\item[\it i)] when $g^2 < 32 \pi^2 \nu^2$, the off-diagonal propagator has a physical mode with real positive mass $m^2_+$. \\\\
It is also worth observing that, for the particular value $g=16 \pi^2 \nu^2$, corresponding to $\tau=0$, the unphysical mode in the decomposition \eqref{NPoff_f} disappears, as ${\cal R}_{-}=0=m^2_-$. Thus, for that particular value of the gauge coupling, the off-diagonal propagator reduces to a single physical Yukawa mode with mass $16\pi^2\nu^4$, {\it i.e.}
\begin{equation}
\left\langle A_{\mu }^{\alpha }(q)A_{\nu }^{\beta }(-q)\right\rangle
= \delta ^{\alpha \beta }  \left(  \frac{1}{q^2+16\pi^2\nu^4} \right)
\left( \delta _{\mu\nu }-\frac{q_{\mu }q_{\nu }}{q^{2}}\right)  \;,  \label{off-Yuk}
\end{equation}
\item[\it ii)] when $g^2>32\pi^2\nu^2$, corresponding to $\tau> \frac{g^2\nu^2}{4}$, all masses become complex and the off-diagonal propagator becomes of the Gribov type with two complex conjugate poles. This region corresponds to a phase in which all gauge modes are confined.
\end{itemize}
In summary, when the Higgs field is in the adjoint representation, we have two phases. For $g^2<32\pi^2\nu^2$ the $A_3$ mode is confined while the off-diagonal propagator displays a physical Yukawa mode with mass $m^2_+$. This phase is referred to as the $U(1)$ symmetric phase \cite{Nadkarni:1989na,Hart:1996ac}. When $g^2>32\pi^2\nu^2$ all propagators are of the Gribov type, displaying complex conjugate poles. This means that all gauge modes are confined. According to \cite{Nadkarni:1989na,Hart:1996ac}, this phase is called the $SU(2)$ confined phase. Since these results were obtained in a semi-classical (= lowest order in the loop expansion) approximation, let us comment on the applicability of such approximation. The latter is reasonable when the \emph{effective} coupling constant is sufficiently small. This effective coupling will for sure contain, in $3d$, the combination $G^2=\frac{g^2}{(4\pi)^{3/2}}$. This does not make sense yet due to $g^2$ having a mass dimension $1$. In the presence of a mass scale $M$, the perturbative series for e.g.~the gap equation will organize itself automatically in a series in $G^2/M$. Specifically, let us assume that we are in the Higgs phase with thus $g^2 < 32 \pi^2 \nu^2$, the effective coupling will be sufficiently small when $\frac{G^2}{\nu^2}$ is small compared\footnote{The Higgs mass $\nu^2$ is then the only mass scale entering the game.} to $1$. Such condition is not at odds with the retrieved condition $g^2 < 32 \pi^2 \nu^2$. Next, assuming the coupling $g^2$ to get large compared to $\nu^2$, thereby entering the confinement phase with $cc$ masses, $g^2$ dominates everything, leading to a Gribov mass scale $\tau\propto g^8$, and an appropriate power of the latter will secure a small effective expansion parameter consistent with the condition $g^2 > 32 \pi^2 \nu^2$. We thus find that at sufficiently small and large values of $\frac{g^2}{\nu^2}$ our approximation and results are trustworthy.

\section{The case of the fundamental representation}
Let us face  now a Higgs field in the fundamental representation of $SU(2)$. In this case, the Lagrangian of the model is given by
\begin{equation}
S=\int d^{3}x\left(\frac{1}{4}F_{\mu \nu }^{a}F_{\mu \nu }^{a}+
(D_{\mu }^{ij}\Phi ^{j})^{\dagger}( D_{\mu }^{ik}\Phi ^{k})+\frac{\lambda }{2}\left(
\Phi ^{\dagger}\Phi-\nu ^{2}\right) ^{2}+b^{a}\partial _{\mu }A_{\mu
}^{a}+\bar{c}^{a}\partial _{\mu }D_{\mu }^{ab}c^{b}\right)  \;, \label{Sf}
\end{equation}
where the covariant derivative is defined by
\begin{equation}
D_{\mu }^{ij}\Phi^{j} =\partial _{\mu }\Phi^{i} -ig \frac{(\tau^a)^{ij}}{2}A_{\mu }^{a}\Phi^{j} \;.
\end{equation}
The indices $i,j=1,2$ refer to the fundamental representation, and $\tau^a, a=1,2,3$, are the Pauli matrices. In this case, for the {\it vev} of the Higgs field we have
\begin{equation}
\langle \Phi \rangle  = \left( \begin{array}{ccc}
                                          0  \\
                                          \nu
                                          \end{array} \right)  \;,  \label{vevf}
\end{equation}
so that all components of the gauge field acquire the same mass $m^2= \frac{g^2\nu^2}{2}$. In fact, for the quadratic part of the action we have now
\begin{equation}
S_{quad}=\int d^{3}x\left( \frac{1}{4} { \left(  \partial_\mu A^a_\nu -\partial_\nu A^a_\mu  \right)} ^2 + b^a \partial_\mu A^a_\mu
+ \frac{g^{2}\nu^{2}}{4}A_{\mu }^{a}A_{\mu }^{a}  \right)  \;. \label{quadf}
\end{equation}
The implementation of the Gribov region can be done exactly as in the previous section, the only difference being that in the present case only one Gribov parameter is needed, due to the fact that all components of the gauge field have the same mass. This immediately leads to the following gluon propagator
\begin{equation}
\left\langle A_{\mu }^{a}(q)A_{\nu }^{b}(-q)\right\rangle
=\delta^{ab}\frac{ q^2}{q^{4} + \frac{g^{2}\nu^{2}}{2} q^2
+ \frac{4g^2}{9} \vartheta }\left( \delta _{\mu
\nu }-\frac{q_{\mu }q_{\nu }}{q^{2}}\right)  \label{propf} \;,
\end{equation}
where $\vartheta$ is the Gribov parameter, determined by the following gap equation
\begin{equation}
\frac{4}{3}g^2 \int \frac{d^3q}{(2\pi)^3} \frac{1}{ q^{4} + \frac{g^{2}\nu^{2}}{2} q^2
+ \frac{4g^2}{9}\vartheta}  = 1  \;, \label{gapf}
\end{equation}
which gives
\begin{equation}
 \frac{4g^2}{9}\vartheta = \frac{1}{4} \left( \frac{g^2\nu^2}{2} -\frac{g^4}{9\pi^2} \right)^2 \;. \label{solf}
\end{equation}
 We can now proceed to analyze the gluon propagator  \eqref{propf}. As done in the previous section, we decompose it in the following way
\begin{equation}
\left\langle A_{\mu }^{a}(q)A_{\nu }^{b}(-q)\right\rangle
=\delta^{ab}  \left(  \frac{{\cal F}_{+}}{q^2+m^2_+} -\frac{{\cal F}_{-}}{q^2+m^2_-}  \right)
\left( \delta _{\mu\nu }-\frac{q_{\mu }q_{\nu }}{q^{2}}\right)  \;,  \label{decf}
\end{equation}
with
\begin{equation}
m^2_+ = \frac{1}{2} \left( \frac{g^2\nu^2}{2}  + \sqrt{\frac{g^6}{9\pi^2} \left(\nu^2-\frac{g^2}{9\pi^2}\right)}\; \right) \;,  \qquad m^2_- = \frac{1}{2} \left( \frac{g^2\nu^2}{2}  - \sqrt{\frac{g^6}{9\pi^2} \left(\nu^2-\frac{g^2}{9\pi^2}\right)} \;\right)
\label{massesf} \;,
\end{equation}
and
\begin{equation}
{\cal F}_{+} = \frac{m^2_+}{m^2_+-m^2_-}  \;, \qquad {\cal F}_{-} = \frac{m^2_-}{m^2_+-m^2_-}  \label{residf} \;.
\end{equation}
Similarly to the previous case, we can distinguish two regions in the $(\nu^2,g^2)$ plane:
\begin{itemize}
\item[\it i)] when $g^2 < 9\pi^2 \nu^2$ both masses $(m^2_+,m^2_-)$ are positive, as well as the residues $( {\cal F}_{+},{\cal F}_{-})$. The gluon propagator, eq.\eqref{decf}, decomposes into two Yukawa modes. However, due to the relative minus sign in expression \eqref{decf} only the heaviest mode with mass $m^2_+$ represents a physical mode. We see thus that, for $g^2 < 9\pi^2 \nu^2$, all components of the gauge field exhibit a physical massive mode with mass $m^2_+$. This region is what can be called a Higgs phase. \\\\Also, let us notice that, for the particular value $g^2=\frac{9\pi^2}{2}\nu^2$, corresponding to a vanishing Gribov parameter $\vartheta=0$, the unphysical Yukawa mode in expression  \eqref{decf} disappears, as $m^2_-= {\cal F}_{-}=0$. As a consequence, the gluon propagator reduces to that of a single physical mode, namely
\begin{equation}
\left\langle A_{\mu }^{a}(q)A_{\nu }^{b}(-q)\right\rangle
=\delta^{ab}   \left( \delta _{\mu\nu }-\frac{q_{\mu }q_{\nu }}{q^{2}}\right) \frac{1}{q^2 +\frac{9\pi^2}{4}\nu^4 } \;.  \label{decff}
\end{equation}
\item[\it ii)] when $g^2> 9 \pi^2 \nu^2$, the masses $(m^2_+,m^2_-)$ become complex. In this region, the gluon propagator, eq.\eqref{decf}, becomes of the Gribov type, displaying complex conjugate poles. All components of the gauge field become thus unphysical. This region corresponds to the confining phase.
\end{itemize}
Summarizing, when the Higgs field is in the fundamental representation, a Higgs phase is detected for $g^2 < 9\pi^2 \nu^2$. When $g^2> 9 \pi^2 \nu^2$, the confining phase emerges. Concerning the trustworthiness of the results, completely analogous comments as in the adjoint case apply here as well.

\section{Conclusion}
In this work the dynamics of  $3d$ Yang-Mills theories in presence of Higgs fields has been investigated from the point of view of the Gribov issue, {\it i.e.} by taking into account the existence of the Gribov copies. As a consequence of the restriction of the domain of integration in the functional integral to the Gribov region $\Omega$, the propagator of the gluon field gets deeply modified by the presence of the non-perturbative Gribov parameters as well as of the {\it vev} of the Higgs field. Looking thus at the structure of the poles of the propagator, we are able to distinguish different regions in the $(g^2,\nu^2)$ plane for the physical spectrum of the theory. Both adjoint and fundamental representation for the Higgs field have been considered, leading to a quite different spectrum. \\\\In the case of the adjoint representation, it turns out that the $A^3_\mu$ mode of the gauge field is always confined. Its propagator is of the Gribov type, exhibiting two unphysical complex conjugate masses. For the off-diagonal components, we can have a different behavior, according to the size of the coupling constant $g$. When $g^2 < 32 \pi^2 \nu^2$, the off-diagonal components display  a physical Yukawa mode, while when $g^2 > 32 \pi^2 \nu^2$ they become confined. According to \cite{Nadkarni:1989na}, these regions have been called the $U(1)$ symmetryic region and the $SU(2)$ confined region, respectively. \\\\The situation turns out to be  different in the case of the fundamental representation. Here, for   $g^2 < 9\pi^2 \nu^2$, all components of the gauge field display a physical Yukawa mode with mass $m^2_+$, eq.\eqref{massesf}. This is what can be called the Higgs phase. Moreover, when $g^2 > 9\pi^2 \nu^2$,  the gluon propagator becomes of the Gribov  type with complex conjugate masses. This is the confining phase. \\\\It is worth to point out that the poles of the gluon propagator are continuous functions of the parameters $(g^2,\nu^2)$. In this sense, the two regions, {\it i.e.} the Higgs and the confining phases, can be seen as being smoothly connected. Let us also underline that, in the present investigation, the quartic self-coupling of the Higgs field $\lambda$ has been implicitly taken to be very large, so that the modulus of the Higgs field gets frozen to its classical {\it vev}.  It is remarkable that our results  are in good agreement with both theoretical results \cite{Polyakov:1976fu} as well as with the data of lattice numerical simulations \cite{Nadkarni:1989na,Hart:1996ac} at sufficiently large values of the Higgs quartic self-coupling $\lambda$. To some extent, this can be regarded as an important test of the nonperturbative study of Yang-Mills theories from the point of view of the Gribov issue, {\it i.e.} by taking into account the presence of the Gribov horizon. \\\\Evidently, some questions remain. Beyond the here adopted semi-classical approximation, we would need to investigate more closely the role of the wrong-signed Yukawa mode in the Higgs phase. Perhaps the clue will be that its residue is in absolute value always smaller than that of the physical mode. In particular, there is a particular value of the coupling where the ``ghost Yukawa'' vanishes. \\\\In summary, the present work aims at giving a contribution to the complex and physically relevant issue of the transition between the Higgs and the confining phases. The results which we have obtained so far are in agreement  with previous findings and can be regarded as a promising step towards the study of more realistic theories such as a $4d$ gauge theory with gauge group  $SU(2) \times U(1)$\cite{prep}.

%%%%%%%%%%%%%%%%%%%%%%%%%%
\section*{Acknowledgments}
%%%%%%%%%%%%%%%%%%%%%%%%%%%
The Conselho Nacional de Desenvolvimento Cient\'{\i}fico e
Tecnol\'{o}gico (CNPq-Brazil), the Faperj, Funda{\c{c}}{\~{a}}o de
Amparo {\`{a}} Pesquisa do Estado do Rio de Janeiro, the Latin
American Center for Physics (CLAF), the SR2-UERJ,  the
Coordena{\c{c}}{\~{a}}o de Aperfei{\c{c}}oamento de Pessoal de
N{\'{\i}}vel Superior (CAPES)  are gratefully acknowledged. D.~D.~is supported by the Research-Foundation Flanders.

\end{document}